\documentclass[amsmath,amssymb,amsbsy,prl,twocolumn,superscriptaddress,showpacs]{revtex4}
\usepackage{amsfonts}
\usepackage{amsmath}
\usepackage{amssymb}
\usepackage{bm}
\usepackage[usenames, dvipsnames]{color}
\usepackage{dcolumn}
\usepackage{epsfig}
\usepackage{eucal}
\usepackage{graphicx}
\usepackage[breaklinks,colorlinks = true,linkcolor = red,urlcolor=cyan,citecolor=red]{hyperref}
\usepackage{mathrsfs}
\usepackage{subfigure}
\usepackage{textcomp}
\usepackage{times}
\usepackage{xy}
\usepackage{color}

\usepackage[toc,page]{appendix}

\begin{document}

\title{Prediction on Properties of Rare-earth 2-17-X Magnets Ce$_2$Fe$_{17-x}$Co$_x$CN : A Combined Machine-learning and Ab-initio Study}
\author{Anita Halder}
\affiliation{Department of Condensed Matter Physics and Material Sciences, S. N. Bose National Centre for Basic Sciences, JD Block, Sector III, Salt Lake, Kolkata, West Bengal 700106, India.}
\author{Samir Rom}
\affiliation{Department of Condensed Matter Physics and Material Sciences, S. N. Bose National Centre for Basic Sciences, JD Block, Sector III, Salt Lake, Kolkata, West Bengal 700106, India.}
\author{Aishwaryo Ghosh}
\affiliation{Department of Physics, Presidency University, Kolkata 700073, India}
\author{Tanusri Saha-Dasgupta}
\email{t.sahadasgupta@gmail.com}
\affiliation{Department of Condensed Matter Physics and Material Sciences, S. N. Bose National Centre for Basic Sciences, JD Block, Sector III, Salt Lake, Kolkata, West Bengal 700106, India.}

\date{\today}

\begin{abstract}
 We employ a combination of machine learning and first-principles calculations to predict magnetic
 properties of rare-earth lean magnets. For this purpose, based on training set constructed out of
 experimental data, the machine is trained to make predictions on magnetic transition temperature
 (T$_c$), largeness of saturation magnetization ($\mu_0$M$_s$), and nature of the
 magnetocrystalline anisotropy (\textcolor{black}{K$_u$}). Subsequently, the quantitative values of $\mu_0$M$_s$ and
 \textcolor{black}{K$_u$} of the yet-to-be synthesized compounds, screened by machine learning, are calculated by
 first-principles density functional theory. The applicability of the proposed technique of combined
 machine learning and first-principles calculations is {demonstrated} on 2-17-X magnets,
 Ce$_2$Fe$_{17-x}$Co$_x$CN. Further to this study, we explore stability of the proposed
 compounds by calculating vacancy formation energy of small atom interstitials (N/C).
 Our study indicates a number of compounds in the proposed family, offers the
 possibility to become solution of cheap, and efficient permanent magnet.
\end{abstract}

\maketitle
\noindent
\section{Introduction}

\begin{figure*}
\includegraphics[width=0.8\linewidth]{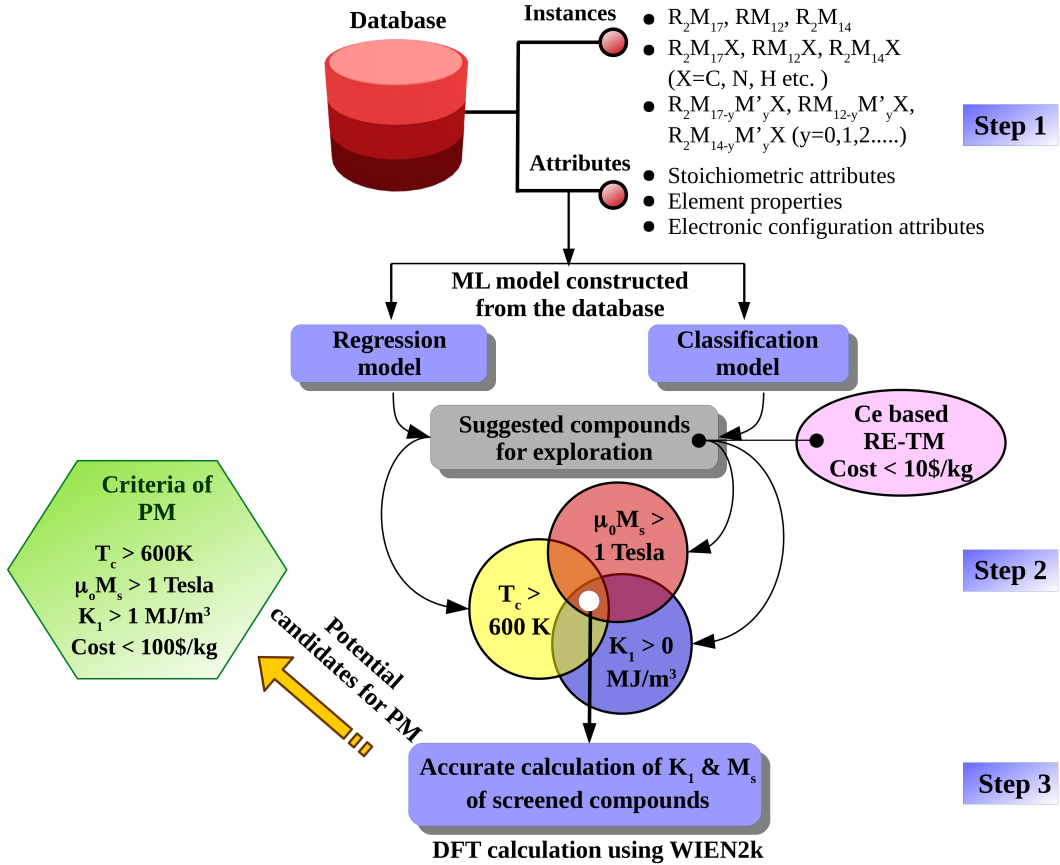}
\caption{(Color online) Steps of Machine learning combined DFT approach for predictions of properties in Ce$_2$Fe$_{17-x}$Co$_x$CN permanent magnets. }
\end{figure*}

\textcolor{black}{Permanent magnets are a part of almost all the most important technologies, 
starting from acoustic transducers, motors and generators, magnetic field and imaging systems to more recent technologies like computer hard disk drives, medical equipment, magneto-mechanics etc.\cite{PM}} The search for efficient permanent magnets is thus everlasting. In this connection, the family of rare-earth (RE) and {{\it 3d}} transition metal (TM) based intermetallics has evolved over last 50 years or so, and has transformed the landscape of permanent magnets.\cite{RE-TM,sun} Two most prominent examples of RE-TM permanent magnets, that are currently in commercial production, together with hard magnetic ferrites, are SmCo$_5$, and NdFe$_{14}$B.

While SmCo$_5$ and NdFe$_{14}$B provide reasonably good solutions, keeping in mind the resource criticality of RE elements like Nd and Sm, a significant amount of effort has been put forward in search of new permanent magnets without critical RE elements or with less content of those. The idea is to optimize the price-to-performance ratio.\cite{RE-TM} This has lead to two routes, (a) search for potential magnets devoid of rare-earth elements,\cite{RE-free} and (b) designing of rare-earth lean intermetallics using abundant RE elements such as La and Ce instead of Sm and Nd.\cite{acta,Pandey,Ce} As stressed by Coey,\cite{coey2012} the demand in hand is to seek for new, low-cost magnets with maximum energy product bridging the ferrites and presently used RE magnets. Following the route (b), cheap, new ternary and quartnary RE-lean RE-TM intermetallics need to be explored, as binaries have been well explored. In parallel, Co being expensive, it may be worthwhile to focus on intermetallic compounds containing Fe.

Starting from the simplest binary RE-TM structure of CaCu$_5$, by replacing $n$ out of $m$ RE (R) sites with a pair of TM (M) sites, R$_{m-n}$M$_{5m+2n}$ structures are obtained. This can give rise to several possible binary structures of different chemical compositions, listed in order of RE-leanness; RM$_{13}$ (7.1$\%$), RM$_{12}$ (7.7$\%$), R$_2$M$_{17}$ (10.5$\%$), R$_2$M$_{14}$ (12.5 $\%$), RM$_5$ (16.7$\%$), R$_6$M$_{23}$ (20.7 $\%$), R$_2$M$_7$ (22.2 $\%$), RM$_3$ (25 $\%$), RM$_2$ (33 $\%$) etc. Judging by the rare-earth content, 1:13, 1:12, 2:17, 2:14 compounds may form examples of rare-earth lean materials. It is desirable to modify the known binary compounds containing low cost RE's belonging to these families to achieve best possible intrinsic magnetic properties, namely (i) high spontaneous or saturation magnetization ($\mu_0$M$_s$), at least around 1T, (ii) a Curie temperature (T$_c$) high enough for the contemplated devise use, 600 K or above, and (iii) a mechanism for creating sufficiently high easy-axis coercivity (\textcolor{black}{K$_u$}). The synthesis and optimization of properties of real materials in experiment is both time-consuming and costly, being mostly based on trial and error. Computational approach in this connection is of natural interest to screen compounds, before they can be suggested and tested in laboratory. Typical computational approaches in this regard are based on density functional theory (DFT) calculations. A detailed calculation estimating all required magnetic properties, {\it i.e} M$_s$, T$_c$, \textcolor{black}{K$_u$} from first-principles is expensive and also not devoid of shortcomings. For example, estimation of T$_c$ relies on parametrization of DFT or supplemented $U$ corrected theory of DFT+$U$ total energies to construct spin Hamiltonian and solution of spin Hamiltonian by mean field or Monte Carlo method. While this approach would work for localized insulators, its application to metallic systems with itinerant magnetism is questionable, as it fails even for elemental metals like Fe, Co and Ni.\cite{fe-co-ni} A more reasonable approach of DFT+dynamical mean field (DMFT)\cite{dmft} is significantly more expensive. An alternative approach would be to use machine learning (ML) technique based on a suitable training dataset. This approach has been used for RE-TM permanent magnets based on DFT calculated magnetic properties database of M$_s$ and \textcolor{black}{K$_u$}.\cite{acta,srep} Creation of database based on calculations, even with high throughput calculations is expensive, and relies on the approximations of the theory. It would be far more desirable to built a dataset based on experimental results, and then train the ML algorithm based on that. However, the size and availability of the experimental data in required format can be a concern. Focusing on the available experimental data on RE lean intermetallics, the set of T$_c$ is largest, followed by that for \textcolor{black}{K$_u$}, and M$_s$. While the quantitative values of T$_c$'s in Kelvin or degree Celsius are available in literature, for magnetocrystalline anisotropy often only the information whether they are easy-axis or easy-plane
are available. Similarly, the $\mu_{o}$M$_s$ values are reported either in $\mu_B/f.u.$ or in emu/gm or in Tesla, conversion from $\mu_B/f.u.$ and emu/gm to
Tesla requiring information of the volume and density, which may introduce inaccuracies up to one decimal point. Restricting experimental data to those containing values of \textcolor{black}{K$_u$},
and $\mu_{o}$M$_s$ values in the same format (either Tesla or $\mu_B/f.u.$ or emu/gm) reduces
the dataset of \textcolor{black}{K$_u$} and M$_s$ significantly, making application of ML questionable. We thus use a two-prong approach, as illustrated in Fig. 1. We first create a database of T$_c$, M$_s$ and \textcolor{black}{K$_u$} from available experimental data on RE-lean intermetallics, and use ML
for prediction of T$_c$ values, for predicting whether $\mu_0$M$_s$ satisfies the criteria of being larger than 1 {Tesla}, and for predicting the sign of \textcolor{black}{K$_u$}. For M$_s$ and \textcolor{black}{K$_u$}, ML thus serves the purpose of initial screening.
We next evaluate M$_s$ and the magnetic anisotropy properties based on elaborate DFT calculations. Calculation of the
magnetic anisotropy energy (MAE) is challenging due to its extremely small value. However, since the pioneering work of Brooks,\cite{brooks} several studies\cite{mazin,smco5,Pandey,durga} have shown that $U$ corrected DFT generally reproduces the orientation and the right order of magnitude of the MAE.\\
\indent
We demonstrate applicability of our proposed approach on Ce and Fe based 2:17 RE-TM intermetallics, Ce$_2$Fe$_{17-x}$Co$_{x}$ compounds ($x$ = 1, $\ldots$, 7).
\textcolor{black}{Our choice is based on following criteria,
 (a) the compounds contain rare earth Ce which is the cheapest one among the RE family having market
  price of $\sim$ 5 USD/Kg.\cite{marketprice} The cost of other components Fe, C and N are all $<$ 1 USD/Kg.
  The price of Co is higher than Fe,\cite{marketprice} being less abundant metal. The Co:Fe ratio is thus restricted within
  0.4.
\textcolor{black}{(b) Co substitution in place of Fe has been reported\cite{odkhuu1,odkhuu2} to be efficient in simultaneous enhancements of K$_u$ as well as
T$_c$ in several TM magnets. This is in sharp contrast to other TM substitutes, such as Ti, Mo, Cr, and V, where magnetic anisotropy as well as
T$_c$ are generally suppressed.}
(c) the search space belongs to 2:17 family, which is the family in which most of the instances in our training set belongs to.
(d) this class of compounds is found to be more stable than the well explored 1:12 compounds.
(e) for large saturation magnetization it is desirable to use Fe-rich compounds, which is also less expensive compared to Co.
(f) although Ce has negative second order Stefan's factor which favors in-plane MAE, experimental findings support that the nitrogenation and carbonation can switch the MAE from easy plane to easy axis.\cite{xu}
(g) though R$_2$Fe$_{17}$ compounds display large magnetization value due to high Fe content, these compounds are disadvantageous as they exhibit low Curie temperature.\cite{book} Presence of Co, as well as C/N interstitials help in increasing T$_c$.
(h) while magnetic properties of carbo-nitrides are expected to be similar to that of nitrides for sufficiently high concentration of N, carbo-nitride compounds have been proven to show better thermal stability.\cite{chen}}

Our study suggests that Fe-rich Ce$_2$Fe$_{17-x}$Co$_{x}$CN compounds may form potential candidate materials for low-cost permanent magnets, \textcolor{black}{satisfying the necessary requirements of a permanent magnet with T$_c$ $>$ 600 K, $\mu_0$M$_s$ $>$ 1 Tesla and easy-axis \textcolor{black}{K$_u$} $>$ 1 MJ/m$^{3}$. The calculated maximal energy product and estimated anisotropy field, which are technologically interesting figures of merit for hard-magnetic materials, turn to be within the reasonable range. Some of the studied compounds may possibly
bridge the gap between low maximal energy product and high anisotropy field for SmCo$_5$ and vice versa for Nd$_2$Fe$_{14}$B.}

\section{Machine learning approach}

\subsection{\textcolor{black}{Database construction \& Training of Model}}

Aiming to search new candidates for permanent magnets we use supervised machine learning (ML) algorithm which helps us to screen compounds with high T$_{c}$ (T$_{c}$ $\gtrsim$ 600 K), high M$_{s}$ ($\mu_0$M$_s>$ 1 Tesla), and easy axis anisotropy ({\textcolor{black}{K$_u$}} $>$ 0) among the huge number of possible candidates of unexplored RE-TM intermetallics. The first step of any ML algorithm is to construct a dataset. We construct three datasets of existing RE-TM compounds for T$_{c}$, M$_{s}$ and \textcolor{black}{K$_u$} separately using the following sources: ICSD,\cite{icsd} the handbook of magnetic materials,\cite{mm} the book of magnetism and magnetic materials,\cite{mmm} and other relevant references.{\cite{debref1, debref2, debref3, debref4, debref5, debref6, debref7, debref8, debref, debref9, debref10, debref11, debref12, debref13, debref14, debref15, k1-1, k1-2, k1-3, k1-5, k1-6, k1-7, k1-9, k1-10, k1-11, k1-12, k1-13, k1-14, k1-15, k1-16, k1-19, k1-20, k1-21, xu, k1-23, k1-24, k1-25, k1-26, k1-27, k1-28, k1-29, k1-31, k1-32, k1-33, k1-34, k1-35, chen,  k1-37, k1-38, k1-39, n1, n2, n3, n4, n5, n6} The datasets are presented as supplementary materials (SM)\cite{suppl} as easy reference for future users. To construct the database of rare-earth lean compounds, RE percentage in the intermetallic compounds is restricted to $14\%$ which includes the four different binary RE-TM combinations namely RM$_{12}$, RM$_{13}$, R$_{2}$M$_{17}$ and R$_{2}$M$_{14}$ along with their interstitial and derived compounds. We discard RM$_{13}$ from the dataset as only few candidates are available from this series with known experimental T$_{c}$, M$_{s}$ and \textcolor{black}{K$_u$}.

We list a total of 565 compounds with reported experimental T$_{c}$, among which majority of the compounds (about $55\%$) belong to R$_{2}$M$_{17}$ series. The minimum contribution to the dataset comes from R$_{2}$M$_{14}$ (about $10\%$) family. The highest T$_{c}$ in the dataset belongs to R$_{2}$M$_{17}$ class of compounds namely Lu$_{2}$Co$_{17}$ \cite{debref1} with T$_{c}\sim$ 1203 K and the compound with lowest T$_{c}$ is {NdCo$_{7.2}$Mn$_{4.8}$ ($\sim$ 120 K),\cite{mm} a member from RM$_{12}$ family. In the dataset all three compositions with RE to TM ratio 2:17, 2:14 and 1:12 show a large variation in T$_{c}$ having the difference between maximum and minimum values as 1051, 775 and 991 K respectively.} There exists few compounds in the dataset with more than one reported value of T$_{c}$. For example T$_{c}$ of SmFe$_{10}$Mo$_2$ has been reported with two different values of 421 K\cite{smfe10mo21} and 483 K.{\cite{smfe19mo22} There are other examples of such multiple T$_{c}$.\cite{mtc1,mtc2,mtc3,mtc4,mtc5} The quality of the sample, their growth conditions, coexistence of compounds in two or multiple phases and accuracy of the measurements may lead to the multiple values of T$_{c}$ reported for a particular compound. In such cases, we consistently consider the largest among the reported values of T$_{c}$. Notably in majority of cases we find little variation in reported values of T$_{c}$ ($\sim$ 20-50 K). 
\indent 

\begin{table}
\begin{tabular}{|c|c|c|c|}
\hline
Attribute Type & Attribute & Notation & Value range \\ 
\hline
Stoichiometric & CW absolute deviation & $<\Delta Z>$ & 1.70-16.74  \\
& of atomic no. & &  \\
& CW av. of & $<Z_{TM}>$ & 10-33.30  \\
& atomic no. of TM &      &            \\
& CW av. of  &  $<Z_{LE}>$ &  0-9.79  \\
& atomic no. of LE &    &         \\
& CW av. Z & $<Z>$ & 21.08-37.71  \\
& CW electronegativity  &  $\Delta \epsilon$  & 0.61-1.84	\\ 
& diff. of RE $\&$ TM & &   \\
& CW RE percentage	& $RE\%$	&	4.76-14.29	\\ 
& CW TM percentage	& $TM\%$	&	38.46-95.24	\\
& CW LE percentage	& $LE\%$	&	0-53.85	\\
  \hline 
Element & Atomic no. of RE  & $Z_{RE}$ & 58-71\\
& Presence of  &  $N_{TM}$	&	yes/no	\\
& more than one TM &            &               \\
& Presence of LE &  $N_{LE}$	&	yes/no	\\
 \hline 
Electronic & Total no. of f electrons & $f^n$ & 1-28 \\ 
& Total no. of f electrons & $d^n$ & 30-136 \\ 
  \hline 

\end{tabular}
\caption{List of 13 different attributes with description, notation and range used in the ML algorithm. Here "CW" stands for "composition-weighted".} 
\end{table}
The dataset of  M$_{s}$ is relatively smaller than T$_{c}$, containing only 195 entries. The majority of the compounds in this dataset belong to 2:17 composition similar to the database of T$_{c}$. The relatively smaller dimension of M$_s$ dataset is primarily due to fact that experimental reports available for M$_s$ are much less than T$_{c}$. Secondly M$_s$ has been mostly reported at room temperature, in some cases at low temperature. To maintain uniformity of the dataset we consider M$_s$ reported at room temperature, resulting in a
lesser number of compounds in the M$_s$ dataset.

Reports with quoted values of anisotropy constant are even more rare. Our exhaustive search resulted in
only 73 data points. This pushes the dataset size to the limit of ML algorithms, for which predictive
capability becomes questionable due to large bias masking the small variance.\cite{njp} On other hand,
if we allow for also experimental data reporting only sign of \textcolor{black}{K$_u$}, this dataset gets expanded to a
reasonable size of 258.\\

After constructing the dataset, we carry out preprocessing of the data, as outlined in Ref.\cite{ourml}. It comprises of removal of noisy data, outliers and correlated attributes. \textcolor{black}{For details see Appendix.}

The next and the most crucial step is to construct a set of simple attributes, which are capable {of describing} the instances (in this case RE-TM compounds) and then deploy ML algorithm to map {them to} a target (in this case T$_{c}$, M$_s$ and \textcolor{black}{K$_u$}).
The attributes considered in this study are summarized in Table. {I}, which can be divided into three broad categories, namely, stoichiometric attributes, element properties and electronic configuration attributes. The stoichiometric attributes may contain the information of both elemental and compositional properties as suggested by Ward et al.\cite{ward} This is based on taking compositional weights (CW) of elemental properties.

In the third step, we train different popular machine learning algorithms with the constructed dataset for prediction. We use ML algorithm in three different problems; (a) to predict the compounds with T$_c$ more than 600 K, (b) compounds with $\mu_0$M$_s >$ 1 Tesla, and (c) compounds with easy-axis anisotropy. Regression is used in the former case, whereas latter two cases are treated as classification problems. We use five different ML algorithms for regression in case of T$_c$ namely Ridge Regression (RR),\cite{RR} Kernel Ridge Regression (KRR),\cite{KRR} Random Forest (RF),\cite{RF, RF1} Support Vector Regression (SVR)\cite{SVR} and Artificial Neural Network (ANN).\cite{ANN} \textcolor{black}{The details can be found in Appendix.}
Out of the five different ML algorithms, it is seen that random forest performs best, which has been also successfully used for prediction of Heusler compounds,\cite{heusler} half-Hausler compounds,\cite{hh} double perovskite compounds,\cite{ourml} half-Heusler semiconductor with low-thermal-conductivity,\cite{half-heusler} zeolite crystal structure classification\cite{zeolite} etc. \textcolor{black}{Results presented in the following are based on random forest method.}

\subsection{Model evaluation}
\begin{figure}[h]
\includegraphics[width=1.05\linewidth]{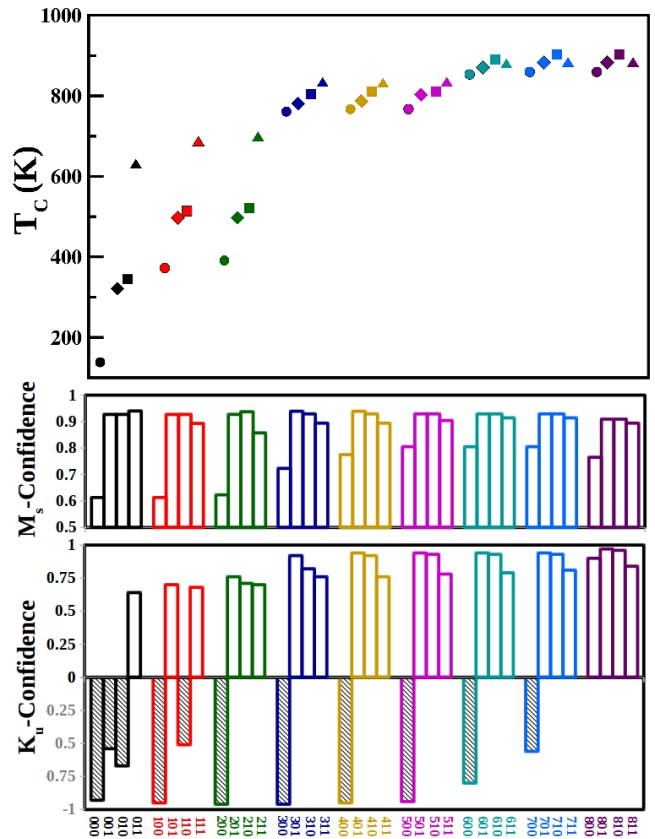}
\caption{(Color online) ML predictions of Curie temperature (T$_c$) from regression model, and saturation magnetization (M$_s$) and anisotropy constant (\textcolor{black}{K$_u$}) from classification model. The upper (middle/lower) panel shows the results of T$_c$ (M$_s$/\textcolor{black}{K$_u$}). The exploration set is Ce$_2$Fe$_{17-x}$Co$_{x}$C$_y$N$_z$ where $y$ and $z$ can have values either {0 or 1}, and $x$ = 0 $\ldots$ 8, acronymed as $x$$y$$z$. In the top panel, non-interstitial compounds, carbonated, nitrogenated and carbo-nitrogenated compounds are symbolized by circle, diamond, square and upper triangle.
Different colors specify compounds with different $x$ values. The middle panel shows the ML prediction
confidence for M$_s$. In the lower panel, ML prediction confidence for \textcolor{black}{K$_u$} is illustrated. Here the upper (lower) half having bars with no-fill (shaded) shows the confidence for the compounds with positive (negative) \textcolor{black}{K$_u$}.}
\end{figure}

\begin{figure*}
  \includegraphics[width=0.7\linewidth]{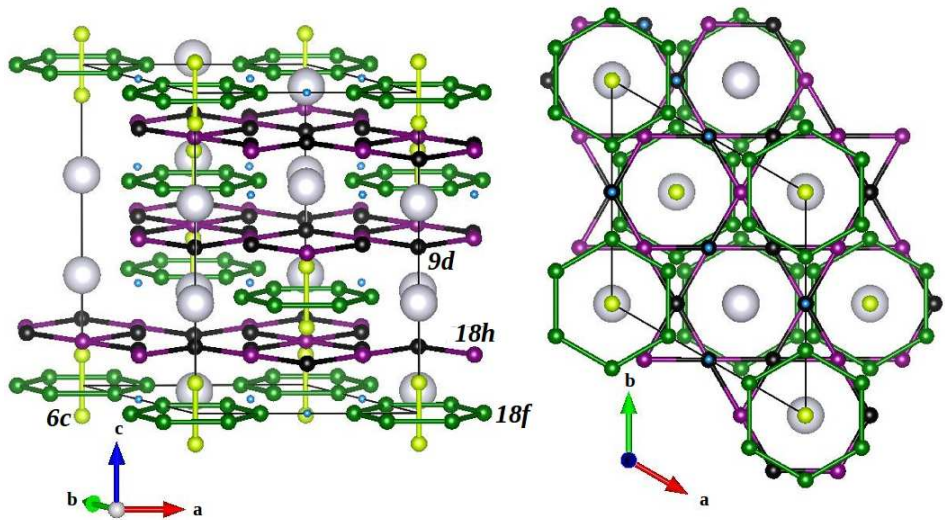}
\caption{(Color online) Crystal structure of Ce$_2$Fe$_{17-x}$Co$_x$CN magnets. The Ce, Fe/Co and C/N atoms are shown with large, medium and small balls, respectively. Four transition metal sublattices 9$d$, 18$f$, 18$h$ and 6$c$ are shown in black, green, magenta and yellow colored balls, respectively. Left panel shows the crystal structure viewed with c-axis pointed vertically up and the right panel shows the crystal structure viewed along the c-axis.}
\end{figure*}

The final step is to employ the trained algorithm on yet-to-be synthesized RE-TM compounds, and thus to explore new compositions with targeted properties. We choose Ce$_2$Fe$_{17-x}$Co$_{x}$C$_y$N$_z$ ($y$,$z$ = 0/1; $x$ = 0 $\ldots$ 8) as the exploration set for application of the trained ML algorithm.
This results in a set of 36 compounds among which 8 compositions (Ce$_2$Fe$_{17-x}$Co$_x$CN, $x$ = 1, $\ldots$, 8) have neither been synthesized experimentally nor studied theoretically, to the best of our knowledge. We apply our trained ML algorithms on all of these 36 compounds and the results are summarized in Fig. 2. The top panel of Fig. 2 shows the predicted T$_c$ of all the compounds. It is seen that the nitrogenation or carbonation increases the T$_c$ with respect to their respective parent compound Ce$_2$Fe$_{17-x}$Co$_{x}$. Our ML model predicts that the nitrides have higher T$_c$ than that of the carbides. For $x$ $\le$ 5, the enhancement of T$_c$ is maximum for the compounds where both carbon and nitrogen are present. For $x$ $>$ 5, T$_c$ shows slight decrease compared to only nitrogenated case. It is also noted that the relative rise in T$_c$ in interstitial compounds compared to parent compounds, decays gradually with Co concentration. The increase in T$_c$ varies from $\sim$ 200 K to 10 K as $x$ varies from 0 to 8 for carbides and nitrides whereas introduction of both nitrogen and carbon shows the variation from $\sim$ 310 K to 30 K. Our result reproduces the trend of experimental findings in a qualitative manner. The experimental results for $x$ = 0 (Ce$_2$Fe$_{17}$),\cite{fuji1992, fuji1995} concluded that the enhancement in T$_c$ is highest in presence of both carbon and nitrogen\cite{alto93,chen93} (T$_c \sim$ 721 K), followed by nitrogenated compound\cite{buschow90, liu91} (T$_c \sim$ 700 K) and lowest for carbonated compound\cite{alto93,chen93} (T$_c \sim$ 589 K). Though it is not possible to compare the results quantitatively as the stoichiometry of the experimentally studied carbonated and nitrogenated compounds are not the same as in our exploration dataset, but the overall trend is similar. We also find that our ML model underestimates the T$_c$ of the pure binary compound Ce$_2$Fe$_{17}$.\cite{book} This is expected, as already discussed, our model is less precise for the prediction of low T$_c$ compounds. \\
\indent
Switching to the M$_s$ part, the middle panel of Fig. 2 shows the confidence of classification of compounds with $\mu_0$M$_s$ more than 1 T. The confidence value closer to 1 implies that the prediction is viable to be more accurate. All the compounds are classified in favor of forming permanent magnets with $\mu_0$M$_s >$1 T. For compounds like Ce$_2$Fe$_{17-x}$Co$_{x}$ the prediction confidence varies from 0.6 to 0.8 with increasing Co concentration, whereas the carbon and nitride compounds are always classified with high prediction
confidence.\\
\indent
The predictions from classification model on \textcolor{black}{K$_u$} is shown in bottom panel of
Fig. 2. We find while the anisotropy of Fe$_{17-x}$Co$_x$ compounds without interstitial C/N ($x$ = 2, $\ldots$ 7) atoms
are predicted to be easy-plane, their carbonated/nitrogenated/carbo-nitrogenated counterparts show
easy-axis anisotropy. For pure Fe compounds, apart from carbo-nitrogenated compound, all are
predicted to be easy-plane, while for Fe$_{16}$Co compounds carbonated as well as carbo-nitrogenated
compounds are predicted to be easy-axis. This in turn, highlights the effectiveness of Co substitution on making \textcolor{black}{K$_u$} positive. We note the prediction
confidence of the carbo-nitrogenated compounds are around 0.75.\\
\indent
On basis of the above ML analysis, we pick up seven yet-to-be synthesized compounds, Ce$_2$Fe$_{17-x}$Co$_x$CN, $x$ = 1, $\ldots$, 7. This choice is guided by the compounds satisfying T$_c >$ 600 K from regression model, and $\mu_0$M$_s >$ 1 Tesla with easy-axis anisotropy from classification models, and being Fe-rich. 
In following, we describe their crystal structure, and present results of DFT calculated electronic structure, anisotropy properties, and stability properties. 
\section{DFT Calculated Properties of Predicted Compounds}

\subsection{Crystal Structure}

The Ce$_2$Fe$_{17}$ compounds crystallize in the rhombohedral Th$_2$Zn$_{17}$-type structure (space group R$\bar{3}$m),
derived from the CaCu$_5$-type structure with a pair (dumbbell) of Fe atoms for each third rare earth atom in the basal plane and
the substituted layers stacked in the sequence ABCABC $\ldots$. As shown in Fig. 3, the transition metal atoms are divided into
four sublattices, 9$d$, 18$f$, 18$h$ and 6$c$, having 3 (9), 6(18), 6 (18), and 2 (6) multiplicity in the one (three) formula unit
primitive-rhombohedral (hexagonal) unit cell.
The TM atoms occupying the 6$c$ sites, referred as dumbbell sites, form the $\ldots$-TM-TM-RE-RE-$\ldots$ chains running along the
c-axis of the hexagonal cell. The 18$f$ TM atoms form a hexagonal layer, which alternates with the hexagonal layer formed by
9$d$ and 18$h$ TM atoms. The 6$c$ TM-TM doumbells pass through the hexagons formed by 18$f$ TM's. For the interstitial C and
N atoms, neutron powder diffraction,\cite{neutron} EXAFS experiments confirmed that they fill voids of nearly octahedral shape
formed by a rectangle of 18$f$ and 18$h$ TM atoms and two RE atoms at opposite corners, which are the 9$e$ sites of  
Th$_2$Zn$_{17}$-type structure, and having the shortest distance from the RE sites among all available interstitial sites.
All our calculations are thus carried out with C/N atoms in 9$e$ positions. 
The RE atoms in 6$c$ position as well as light elements C/N in 9$e$ interstitial sites belong to the same layer as 18$f$ TMs.
As the 9$e$ sites are in the
same $c$-plane with the RE sites, having RE atoms at neighbors, introduction of interstitials like C and N, is expected to have
a profound influence on the the electronic environment of RE atom, thereby altering the magneto-crystalline anisotropy.

Although the R$\bar{3}$m symmetry is lowered upon Co substitution and the spin-orbit coupling (SOC) in the anisotropy calculation,
for the ease of identification, we will still use the the notations 9$d$, 18$f$, 18$h$ and 6$c$. Our total energy calculations show that
Co preferentially occupy sites in the sequence 9$d$ $>$ 18$h$ $>$ 6$c$ $>$ 18$f$. Out of available 17 TM sites we have considered
Co substitution up to 7 sites, which result in Fe-rich phases of compositions Ce$_2$Fe$_{17-x}$Co$_x$CN with $x$ = 1, 2, $\ldots$, 7.
Following the site preference we consider Co atoms in 9$d$ and 18$h$ sites.

\textcolor{black}{We expect the lattice parameters not to change much upon Co substitution, as Fe and Co, being neighboring elements in periodic table, has similar atomic radii. Nevertheless, to check
  the influence of Co substitution on lattice structure, we optimize the lattice constant and the volume for all $x$ values. Following our expectation, the results show only
  a marginal decrease in lattice parameter and volume (with a maximum deviation of 1$\%$) upon increasing Co content, in line with the findings by Odkhuu et al.\cite{odkhuu2} for 1:12 compounds,
  and the experimental findings by Xu and Shaheen on 2:17 compounds.\cite{xu}
This minimal change is found to have no appreciable effect on magnetic properties, as explicitly checked on
representative compounds with $x$ = 1, 4 and 7. We thus choose the
lattice structure as the optimized lattice structure of $x$ = 0 (see Appendix), with lattice constant = 6.59 $\AA$ and angle $\beta$ = 83.3$^{o}$ of the rhombohedral unit cell\cite{kou-cs} in subsequent calculations.}

\subsection{Magnetic Moment and Electronic Structure}

\begin{figure}
\includegraphics[width=1.0\linewidth]{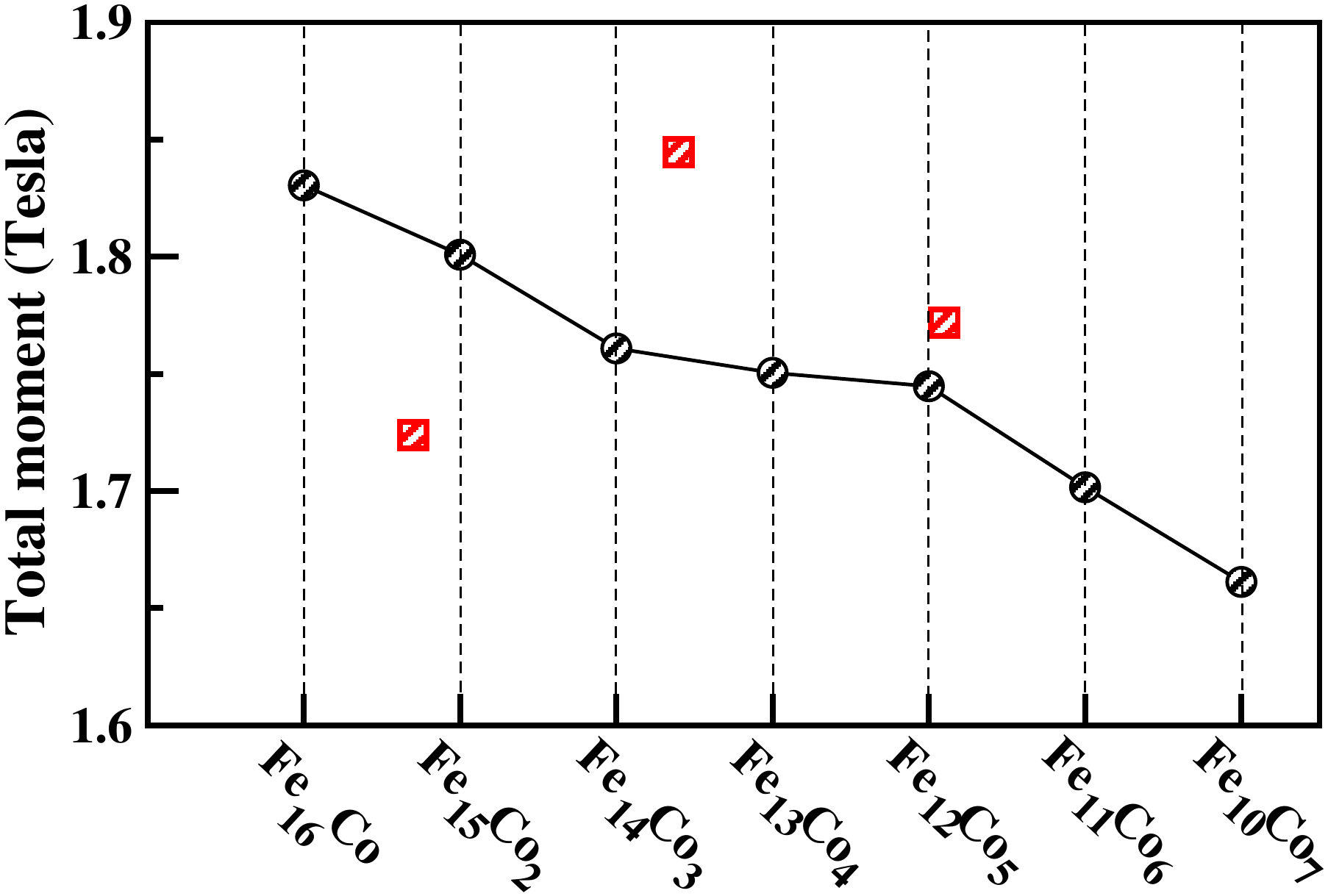}
\caption{(Color online) Calculated total moment (black circles), $\mu_0$M in Tesla plotted for increasing Co concentrations of Ce$_2$Fe$_{17-x}$Co$_x$CN compounds. Shown are also experimental results\cite{xu} (red, square) for Ce$_2$Fe$_{17-x}$Co$_x$N$_y$ compounds measured at room temperature. For comparison between T = 0 K calculated moments, and experimental data measured
at room temperature, the experimental data has been scaled by a factor of 1.3.}
\end{figure}

\begin{figure}
\includegraphics[width=1.0\linewidth]{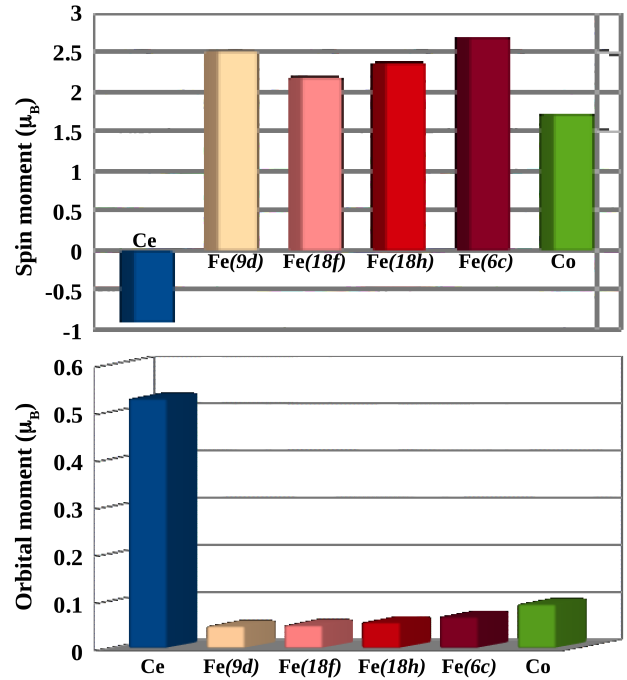}
\caption{(Color online) Calculated spin (top) and orbital (bottom) moments at Ce, Fe(9$d$), Fe(18$f$), Fe(18$h$), Fe(6$c$) and Co
 sites in the representative case of Ce$_2$Fe$_{15}$Co$_2$CN compound.}
\end{figure}

\begin{figure*}
\includegraphics[width=0.7\linewidth]{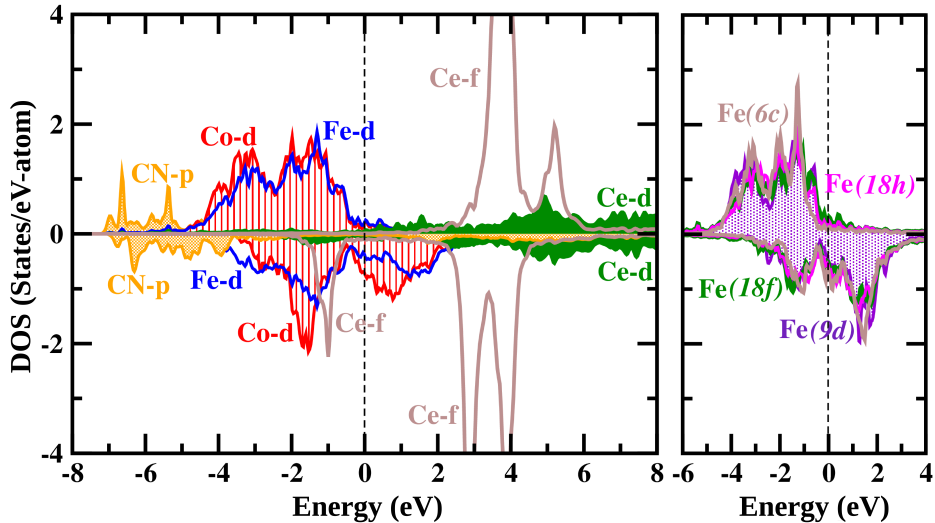}
\caption{(Color online) Left: Density of states of Ce$_2$Fe$_{15}$Co$_2$CN compound, projected onto
 Ce $f$ (brown), Ce $d$ (shaded green), Fe $d$ (blue), Co $d$ (shaded red) and CN $p$ (shaded orange) characters.
 Right: Density of states of Ce$_2$Fe$_{15}$Co$_2$CN compound projected to different Fe $d$'s: Fe(9$d$)
 (shaded indigo), Fe(18$h$) (magenta), Fe(18$f$) (green) and Fe(6$c$) (brown). The zero of the energy
 is set at Fermi energy.}
\end{figure*}

In the following we present the DFT results for the magnetic moments and density of states (DOS), as given in GGA+$U$+SOC calculations.
The details of the DFT calculations are presented in the Appendix. Importance of application of supplemented Hubbard $U$ on RE sites
within LDA or GGA+$U$ formalism is considered as one of the possible means to deal with localized $f$ orbitals of RE ions, and have shown
to provide reasonable description.\cite{mazin,smco5} Previous calculations in compounds containing Ce, showed variation of $U$ within 3 eV to 6 eV,
keeps the results qualitatively same.\cite{Pandey,eriksson} In the following, we present results for $U$ applied on Ce atoms chosen to be 6 eV.

Fig. 4 shows the calculated total magnetic moments of the seven mixed Fe-Co compounds, Ce$_2$Fe$_{17-x}$Co$_x$CN ($x$ = 1, 2, $\ldots$7).
The total magnetic moment shows a decreasing trend with increase of Co concentration, \textcolor{black}{arising from the fact
  that Co moment is smaller that of Fe.}  However, it is reassuring to note that even for
compound with largest Co concentration, Ce$_2$Fe$_{10}$Co$_7$CN, the calculated moment is more than 1.65 Tesla. This is in agreement
with ML prediction, which predicts $\mu_0$M$_s$ of all the considered compounds to be larger than 1 Tesla, though it is to be noted
the ML predictions are made for room temperature moments while the DFT calculated moments are at T = 0 K. The measured values of total moment in corresponding nitrogenated compounds show good comparison (cf Fig. 4) with our calculated moments. \textcolor{black}{In particular, barring the data on x $\approx$ 2, the other two data point show good matching with the trend of theoretical results. We note that the
  experimentally determined moments are for Ce$_2$Fe$_{17−x}$Co$_x$N$_y$ compounds, which contains only N as interstitial atom,
  and the value of $y$ is not mentioned, which may even vary depending on value of $x$.}

Fig. 5 shows the spin and orbital moments projected to Ce, Fe(9$d$), Fe(18$f$), Fe(18$h$), Fe(6$c$) and Co atoms for the representative case
of Ce$_2$Fe$_{15}$Co$_2$CN compound. The results for other Co concentrations are similar. In presence of
large SOC coupling at Ce site, a substantial orbital moment develops, which is oppositely aligned to 
its spin moment following Hund's rule. Considering 3+ nominal valence of Ce, it would be in 4$f^{1}$ state,
with S=1/2 and L=3. While the calculated value of Ce spin moment is close to 1 $\mu_B$ ($\approx$ 0.95 $\mu_B$) in
accordance with nominal S=1/2 state, the orbital moment shows significant quenching with a calculated value
of about 0.5 $\mu_B$. This value of orbital moment is in agreement with DFT calculated values of other Ce containing
RE-TM magnets.\cite{Pandey,chauhan} The 4$f$ electrons are coupled to 5$d$ electrons at Ce site by intra-atomic exchange
interaction, following which their spin moments are aligned in parallel direction. The delocalized 5$d$ electrons
at Ce site, hybridize with Fe/Co 3$d$ electrons, favoring antiparallel alignment of Ce and Fe/Co spins,
as found in Fig. 5. The spin magnetic moment at Fe sites show a distribution, with Fe at 6$c$ site having
largest moment, followed by Fe at 9$d$ and 18$h$ sites while Fe at 18$f$ site shows the lowest moment.
\textcolor{black}{We notice that Fe (6$c$) atoms occupying the dumbbell sites, have less connectivity compared to Fe(9$d$),
  Fe (18$f$) and Fe (18$h$), and thus possess the largest moment, being of most localized character. Among Fe (9$d$), Fe(18$f$), Fe(18$h$) sites Fe (18$f$)
  has smallest moment, driven by the fact that interstitial C and N atoms are in same plane as Fe (18$f$) causing enhanced
  $d$-$p$ hybridization, and reduction in moment.} 
These
spin moments though are larger than that of bulk Fe ($\approx$ 2.2 $\mu_B$). The orbital moment at Fe sites are tiny
($\approx$ 0.05 $\mu_B$). In comparison, Co shows significantly smaller spin moment ($\approx$ 1.7 $\mu_B$) and somewhat larger
orbital moment ($\approx$ 0.1 $\mu_B$), justifying the fall in total moment with increasing concentration of Co.

Fig. 6 shows the density of states of Ce$_2$Fe$_{15}$Co$_2$CN, projected to various orbital characters. The Ce 4$f$ states
are all unoccupied in the majority spin channel, partly occupied in the minority spin channel, in accordance with nominal $f^{1}$ occupancy.
The RE 4$f$ - TM 3$d$ hybridization through empty RE 5$d$ states is visible, making the spin splitting at Fe and Co sites antiparallel to
that of Ce. The C/N $p$ states mostly spanning the energy range -7 eV to -4 eV, show non negligible
mixing with Fe $d$, Co $d$ and Ce characters, justifying their role in influencing the magnetic
properties. Fe $d$ and Co $d$ states span about the same energy range from -4 eV to 2 eV, with
states mostly occupied in the majority spin channel and partially occupied in the minority spin
channel, largely accounting for the metallicity of the compound. Spin splitting of Fe $d$ is
larger than that of Co, being consistent with larger magnetic moment of Fe compared to Co.
Projection to different inequivalent Fe sites (cf right panel of Fig. 6), Fe(9$d$), Fe(18$h$),
Fe(18$f$) and Fe(6$c$) shows that Fe(6$c$) belonging to dumbbell pair is distinct from other Fe sites,
which also exhibit largest magnetic moment among all Fe's.

\subsection{Magneto-crystalline Anisotropy}

\begin{figure}
\includegraphics[width=1.0\linewidth]{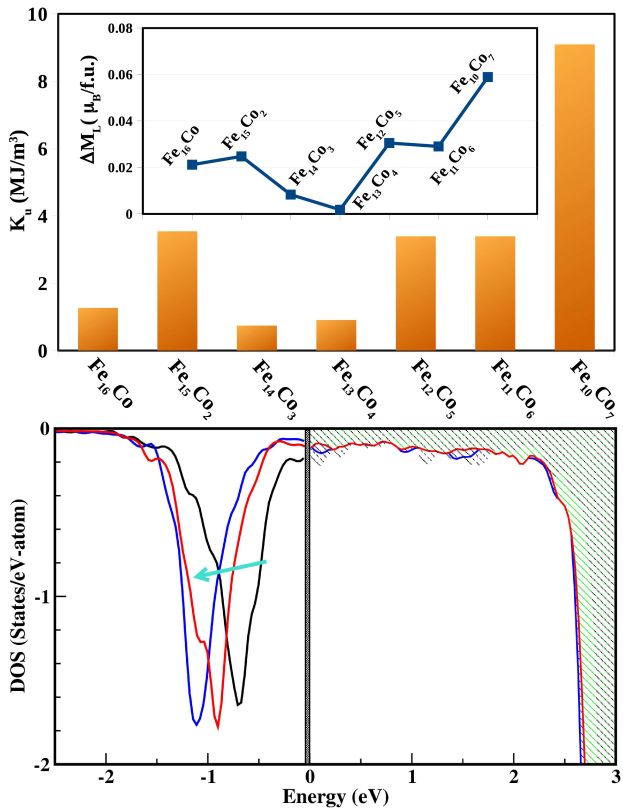}
\caption{(Color online) \textcolor{black}{Top: Calculated magnetocrytalline anisotropy constant in MJ/m$^{3}$ plotted for increasing Co
    concentrations of Ce$_2$Fe$_{17-x}$Co$_x$CN compounds. The inset shows the anisotropy in orbital moment (see text for details). Bottom:
    The GGA+$U$+SOC DOS projected to Ce $f$ energy states with magnetization axis pointed along easy-axis, for Ce$_2$Fe$_{17}$
    (black),  Ce$_2$Fe$_{17}$CN (red) and Ce$_2$Fe$_{16}$CoCN (blue). The zero of the energy is set at Fermi energy, with unoccupied
    part shown shown as shaded. The arrow indicates the shift in occupied part.}}
\end{figure}

\begin{figure*}
\includegraphics[width=0.7\linewidth]{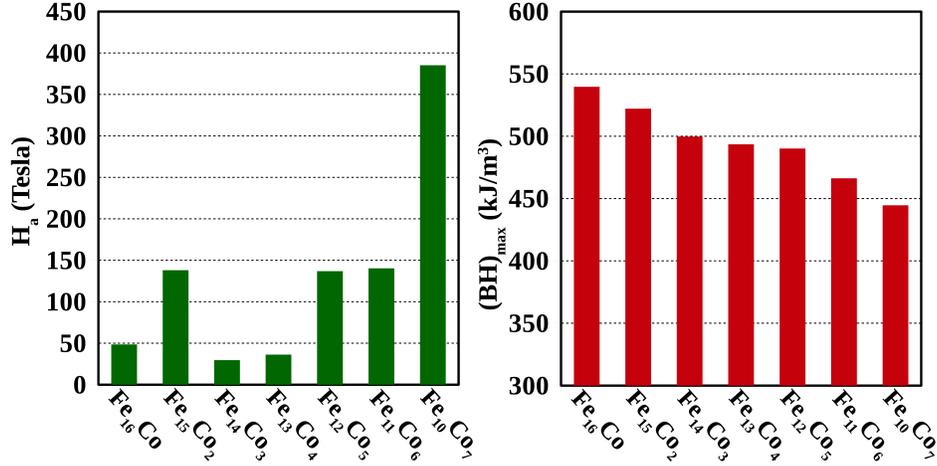}
\caption{\textcolor{black}{(Color online) Calculated anisotropy field in Tesla (left) and maximal energy product in kJ/m$^{3}$ (right) plotted for increasing Co concentrations of Ce$_2$Fe$_{17-x}$Co$_x$CN compounds.}}
\end{figure*}

Having an understanding of the basic electronic structure, in terms of magnetic moments and density of states,
we next focus on calculation of magneto-crystalline anisotropy constant, \textcolor{black}{K$_u$}, which is a crucial quantity responsible
for coercivity in a permanent magnet. \textcolor{black}{MAE defines the energy required for turning the orientation of the magnetic moment under applied field, expressed as $E(\theta) \approx K_1 sin^{2} \theta + K_2 sin^{4} \theta + K_3 sin^{4}\theta cos 4\phi$, where K$_1$, K$_2$, and K$_3$ are the magnetic anisotropy constants, $\theta$ is the polar angle between the magnetization vector and the easy axis ($c$-axis), and $\phi$ is the azimuthal angle
  between the magnetization component projected onto the $ab$ plane and the $a$-axis. In most cases, the higher
  order term K$_3$ is relatively small compared with K$_1$ and K$_2$. For $\theta$ = $\pi$/2, one may thus write
  $K_u \approx K_1 + K_2$.} It's positive and negative values indicate the easy axis and easy plane anisotropy,
respectively. To satisfy the criteria of a good permanent magnet, it should have easy axis anisotropy with value
larger than 1 MJ/m$^{3}$.\cite{coey2012,RE-TM} The MAE in RE-TM arises from two contributions, (i) MAE of the RE sublattice
due to strong spin-orbit coupling and crystal field effect and (ii) MAE of TM sublattice. The interplay of
the two decides the net sign and magnitude. In particular, in the proposed compounds, presence of Co with
significant value of orbital moment, makes the contribution of TM sublattice important.
While 2:17 compounds, primarily show easy plane anisotropy, switching
to easy axis anisotropy for interstitial compounds have been reported. In particular, upon nitrogenation,
easy plane anisotropy has been reported for Ce containing mixed Fe-Co compounds.\cite{xu} As mentioned already,
the interstitial atoms occupy the same plane as the RE atoms, significantly influencing their properties.
With predicted high T$_c$ and large saturation moment of our proposed compounds with carbonation and nitrogenation, it
remains to be seen whether they would exhibit easy axis anisotropy of reasonable values, as required
for a legitimate candidate for permanent magnet. For this purpose, we carry out calculations within
GGA+$U$+SOC with magnetization axis pointing along the crystallographic c-axis and perpendicular to it.
The importance of application of $U$ on proper description of MAE in terms of its sign and order of magnitude has
been stressed upon by several authors.\cite{Pandey,mazin}
In order to establish our method on calculation of MAE involving small energy difference, we first apply
our method to known and well studied case of SmCo$_5$, with choice of $U$ = 6 eV on Sm, and obtained a MAE value of
24.4 meV/f.u, which agrees well with GGA+$U$+SOC calculated value of 21.6 meV/f.u., reported in literature\cite{mazin}
as well as experimentally measured values of 13-16 meV/f.u.\cite{expt}
The calculated results for the proposed Ce$_2$Fe$_{17-x}$Co$_x$CN are shown in top panel of Fig. 7. We
found that MAE shows site-dependence on the Co substitution. We consider configurations with Co
atoms substituting Fe(9$d$) and Fe(18$h$) sites\textcolor{black}{, configurations involving other substituting sites
 being energetically much higher. We consider configurations which are energetically close (within 600 K) and calculate
the Co-composition dependent MAE using the virtual crystal approximation.
Specifically, for $x$ = 1 we consider configurations Co@Fe(9$d$) and Co@Fe(18$h$),
the latter  being 3.58 meV higher compared to former. Similarly for $x$ = 2, we consider  Co@ 2 $\times$ Fe(9$d$) and Co@ 2 $\times$ Fe(18$h$), the latter being 4.43 meV higher compared to former. For $x$ = 3, the configurations considered are, Co@ 2 $\times$ Fe(9$d$)+ Fe(18$h$);  Co@ 3 $\times$ Fe(9$d$);  Co@Fe(9$d$) + 2 $\times$ Fe(18$h$), the energies being 0 meV (set as zero of energy), 12.37 meV and 47.66 meV, respectively. For $x$ = 4, the configurations considered are, Co@ 2 $\times$ Fe(9$d$) + 2 $\times$ Fe (18$h$);  Co@ 3 $\times$ Fe(9$d$) + Fe(18$h$), the energies being 0 meV (set as zero of energy) and 36.5 meV, respectively. For $x$ = 5 , 6 and 7, only one configuration is considered, others being energetically much higher, namely, Co@3 $\times$ Fe(9$d$) + 2 $\times$ Fe(18$h$), Co@3 $\times$ Fe(9$d$)+ 3 $\times$ Fe(18$h$) and Co@3 $\times$ Fe(9$d$) + 4 $\times$ Fe(18$h$), respectively.}

Considering spin-orbit effect
only on Ce atom, it is found to account for about 60$\%$ of the calculated MAE. We find all the calculated
MAE is positive, in good agreement with ML prediction on mixed Fe-Co carbo-nitride compounds. Further MAE
values show non-monotonic dependence on Co concentration. \textcolor{black}{Such non-monotonic trend
  upon varying TM content has been also reported in context of R(Fe$_{1-x}$Co$_x$)$_{11}$TiZ (R = Y and Ce; Z= H, C, and N)\cite{Ce} and R-TM systems in general.\cite{hal} In the inset of top panel of Fig. 7, we show the calculated orbital magnetic anisotropy ($\Delta$M$_L$) defined
  as $\Delta$M$_L$ = M$_L$(a) - M$_L$(c), as employed in Ref.\onlinecite{odkhuu2},  M$_L$(c) and M$_L$(a) being the orbital moment
  along the $c$-axis and $a$-axis, respectively. We find a correlation between $\Delta$M$_L$ and K$_u$, qualitatively satisfying Bruno's expression\cite{bruno} for itinerant ferromagnets given as, K$_u$ = ($\frac{\xi}{4 \mu_B}$) $\Delta$ M$_L$, where $\xi$ is the strength of SOC.}

Most of the easy-axis \textcolor{black}{K$_u$} values are found to be larger than 1 MJ/m$^{3}$, except Fe$_{14}$Co$_3$ and Fe$_{13}$Co$_4$ for which it is found to be 0.74 and 0.91 MJ/m$^{3}$, respectively. Few of the concentrations
exhibit easy-axis \textcolor{black}{K$_u$} values larger than 2 MJ/m$^{3}$,
{\it e.g.} Fe$_{15}$Co$_2$ (3.54 MJ/m$^{3}$), Fe$_{12}$Co$_5$ (3.39 MJ/m$^{3}$), Fe$_{11}$Co$_6$ (3.39 MJ/m$^{3}$),
Fe$_{10}$Co$_{7}$ (9.10 MJ/m$^{3}$), being comparable to Nd$_2$Fe$_{14}$B (4.9 MJ/m$^{3}$).\cite{ndfeb}

\textcolor{black}{To obtain microscopic understanding of the role of Co substitution and doping by C, N on magnetocrystalline anisotropy, we further calculate
the magnetocrystalline anisotropy of Fe-only compounds Ce$_2$Fe$_{17}$, Ce$_2$Fe$_{17}$C, Ce$_2$Fe$_{17}$N and Ce$_2$Fe$_{17}$CN. This results in negative K$_u$ values for Ce$_2$Fe$_{17}$,
  and Ce$_2$Fe$_{17}$C (-2.12 MJ/m$^{3}$ and -1.35 MJ/m$^{3}$), a tiny positive value for Ce$_2$Fe$_{17}$N (0.26 MJ/m$^{3}$) and positive value for co-doped
  compound Ce$_2$Fe$_{17}$CN (1.27 MJ/m$^{3}$). We further plot the GGA+$U$+SOC density of states (cf bottom panel, Fig. 7) with magnetization axis along c-axis projected to Ce $f$ states for Ce$_2$Fe$_{17}$, Ce$_2$Fe$_{17}$CN and Ce$_2$Fe$_{16}$CoCN, which is expected to reveal
  the mechanism of uniaxial anisotropy. We find that a lowering of occupied Ce $f$ energy states and increase in band width occur upon introduction of light elements C and N. This gets further helped by substitution of Co, caused by hybridization between Ce $f$ states and Co $d$ and C,N $p$ states. This gain in hybridization energy stabilizes easy-axis magnetization (cf. Ref.\onlinecite{jsps}) as observed experimentally.\cite{xu}}

\subsection{Maximal energy product and Anisotropy Field}

While, the estimates of {\textcolor{black}{K$_u$}} and $\mu_0$M$_s$ are useful information to access the effectiveness of the suggested
materials as permanent magnets, technologically interesting {figures} of merit of hard magnetic materials, are the
maximal energy product (BH)$_{max}$ and anisotropy field H$_a$. These can be estimated from the knowledge of $\mu_0$M$_s$
and {\textcolor{black}{K$_u$}} as follows,
\[
(BH)_{max} = \frac{(0.9 \mu_0 M_s)^{2}}{4 \mu_0}
\]
\[
H_a = \frac{2 K_u}{\mu_0 M_s}
\]
The factor 0.9 in the expression for (BH)$_{max}$ implies the common assumption that ideally out 10$\%$ of a processed
bulk hard magnet consists of non-magnetic phases.\cite{non} The estimated (BH)$_{max}$ and H$_a$ is shown in Fig. 8.
The (BH)$_{max}$ value is found to range from \textcolor{black}{444 to 540 kJ/m$^{3}$}, in comparison to experimentally measured
values 516 kJ/m$^{3}$ and 219 kJ/m$^{3}$ for Nd$_2$Fe$_{14}$B\cite{en} and SmCo$_5$,\cite{en} respectively. The H$_a$
shows a strong variation with Co concentration, ranging from \textcolor{black}{$\approx$ 1 Tesla to 14 Tesla.}\cite{ha}

We further note that the hardness parameter, defined as $\kappa$ = $\sqrt{\frac{K_u}{\mu_0 M_s^2}}$, turns out to be greater
than 1 for Ce$_2$Fe$_{15}$Co$_2$CN, Ce$_2$Fe$_{12}$Co$_5$CN, Ce$_2$Fe$_{11}$Co$_6$CN, and
Ce$_2$Fe$_{10}$Co$_7$CN compounds, employing the calculated T = 0 K values of \textcolor{black}{K$_u$} and M$_s$.

\subsection{Stability}

\textcolor{black}{Unlike the other RE-TM magnets like 1:12 compounds, one of the advantage of 2:17 compounds is their stability. Both stable form of Ce$_2$Fe$_{17}$ and its Co substituted form have been reported in literature.\cite{xu} Calculation of formation enthalpies, as given in Ref.\onlinecite{odkhuu2}, $E_{form} = \frac{E_{compound}  - \sum_{k} N_{k} \epsilon_{k}}{\sum_{k} N_{k}}$, where $N_{k}$ indicate number of different atoms (Ce, Fe, Co, N and C) in the cell, and $\epsilon_{k}$ denote energy/atom of bulk Ce in FCC structure, $\alpha-$ Fe, Co in HCP structure, in molecular nitrogen and C in graphite structure, gives values -0.61 to -0.59 eV/atom for the studied Ce$_2$Fe$_{17-x}$Co$_x$CN compounds.}

A major challenge with interstitial compounds, though, is the nitrogen diffusion.\cite{chen}
It has been further suggested the blockage of nitrogen
diffusion by carbon layer is useful in reduction of nitrogen outgassing in carbo-nitrides. In particular,
heating up Sm$_2$Fe$_{17}$ carbo-nitrides at a constant rate in a differential scanning calorimeter, the onset
temperature of nitrogen outgassing was found to be higher by more than 40 K, as compared to nitride counterpart.\cite{chen} This justifies the choice of carbo-nitrides as our exploration set. To this end, we
calculate the vacancy formation energy of the interstitial atoms in our chosen compounds. For this
purpose, we calculate the formation energy of the N and/or C vacancy ($\Delta E_{f}$) defined as,
\[
\Delta E_{f} = E^{N(C)_{vac}} - E^{pristine} + E_{N(C)}
\]
where $ E^{N(C)_{vac}}$ and $E^{pristine}$ denote the optimized total energies of compound containing N and/or C vacancy, and vacancy free compound. The internal positions for defect free pristine structure and structures containing nitrogen and/or carbon vacancies are performed keeping the lattice parameters fixed. 
$E_{N(C)}$ is the energy per N or C atom, which is obtained from calculation of N$_2$ molecule
or graphite. The obtained results for Ce$_2$Fe$_{17-x}$Co$_{x}$CN compounds in minimum energy configuration of Co
is shown in Table II. The vacancy formation energies show hardly any variation on chosen configuration for
a given Co concentration.\\
\noindent
\begin{table}
\begin{tabular}{|c|c|c|c|}
\hline
 & $\Delta E_{f} (CN)$ & $\Delta E_{f} (N)$ & $\Delta E_{f} (C)$ \\\hline
$x$ = 1 & 4.32 & 2.10 & 0.97 \\
$x$ = 2 & 3.99 & 2.09 & 0.85 \\
$x$ = 3 & 4.16 & 2.09 & 0.88 \\
$x$ = 4 & 3.98 & 2.10 & 0.79 \\
$x$ = 5 & 3.82 & 2.07 & 0.70 \\
$x$ = 6 & 3.91 & 2.05 & 0.72 \\
$x$ = 7 & 3.78 & 2.01 & 0.69 \\ \hline
\end{tabular}
\caption{\textcolor{black}{Vacancy formation energy for carbon ($\Delta E_{f} (C)$), nitrogen ($\Delta E_{f} (N)$) and
 nitrogen-carbon ($\Delta E_{f} (CN)$) in eV in Ce$_2$Fe$_{17-x}$Co$_{x}$CN compounds.}}
\end{table}
\textcolor{black}{The vacancy formation energies, listed in Table II, show only small variation between compounds of varying
  Co concentration, with the general trend $\Delta E_{f} (CN)$ $>$ $\sum$ ($\Delta E_{f} (N)$ + $\Delta E_{f} (C)$). The individual
  nitrogen vacancy formation energy and carbon vacancy formation energy, are in overall agreement with that found for related compound,
  SmCaFe$_{17}$C(N)$_3$.\cite{Pandey} The vacancy formation energy for co-doped carbon-nitrogen compounds are found to be enhanced by
  about 35-40 $\%$ compared to the sum of the individual C and N vacancy formation energies, proving the carbo-nitrogenation co-doping
  to provide better thermal stability. We also check our results by repeating  vacancy formation energy calculations for $x$ = 0
  compounds, which however do not show significant difference, suggesting Co doping not having major role in stability, as also
  indicated by no significant variation of results between $x$ = 1, 2, 3, 4, 5, 6 and 7.}

\section{Conclusion}

Designing alternative solutions for permanent magnets, satisfying the criteria of low-cost, while keeping
the magnetic properties comparable to those of permanent magnets in use, is of utmost importance for
cost-effective technology. Towards this goal, we use a combined route of machine learning, based
on experimental data, and the first-principles calculations.\textcolor{black}{ While machine learning
  has been applied for problem of rare-earth magnets,\cite{acta} those studies have been based on the
  dataset created out of high throughput calculations. Being dependent on calculation-based inputs,
  creation of such database is not only computationally expensive, but also not devoid of approximations of
  the theory. Our study, to the best of our knowledge, being based on a exhaustive search of experimental data,
  is first of this kind in context of rare-earth magnets.}

While a large volume of experimental
data is available with numerical value of T$_c$, the corresponding dataset with numerical values of
M$_s$ and \textcolor{black}{K$_u$} is small. On the other hand, there exists sizable dataset with information of
\textcolor{black}{K$_u$} being positive (easy axis) or negative (easy plane), and $\mu_0$M$_s$ being larger or smaller
than 1 Tesla. We thus employ regression model of machine learning training to make predictions on
numerical values of T$_c$, and classification model to make predictions on sign of \textcolor{black}{K$_u$}, and $\mu_0$M$_S$
being larger or smaller than 1 Tesla. We apply the trained machine learning to 2:17 rare-earth
transition metal compounds with carbon and nitrogen in interstitials. We choose the compounds to contain
abundant rare-earth Ce, and to be Fe-rich to make them cost-effective. Although nitrogenated version
of this series has been investigated,\cite{xu} the systematic study of the carbo-nitride family to
the best of our knowledge is unavailable. The machine learning predicts T$_c$ of the chosen carbo-nitride
family to be larger than 600 K, $\mu_0$M$_S$ $>$ 1 Tesla, and \textcolor{black}{K$_u$} $>$ 0, thereby indicating the
possibility of them to become good solutions for cost-effective, permanent magnets. Subsequent
first-principles calculations, show T=0 K, $\mu_0$M$_S$ to be larger than 1.65 Tesla, and \textcolor{black}{K$_u$}
$\gtrsim$ 1 MJ/m$^{3}$ for the entire family, Ce$_2$Fe$_{17-x}$Co$_x$CN ($x$ = 1, $\ldots$ 7).
Calculated \textcolor{black}{K$_u$} values are found to be comparable to the state-of-art permanent magnet
Nd$_2$Fe$_{14}$B for Ce$_2$Fe$_{15}$Co$_2$CN, Ce$_2$Fe$_{12}$Co$_5$CN, Ce$_2$Fe$_{11}$Co$_6$CN, and
Ce$_2$Fe$_{10}$Ce$_7$CN. This results in two figure of merits for hard magnets, (BH)$_{max}$ and H$_a$
in range of \textcolor{black}{444-540 kJ/m$^{3}$ and $\approx$ 1 - 14 T, respectively}.

In spite of good magnetic properties, one of the limitation of practical
applications of interstitial 2:17 magnets is the formation of nitrogen/carbon vacancies at high temperature.
\textcolor{black}{By calculating the N-(C)-vacancy formation energy, we show that carbo-nitrogenation co-doping
enhances the vacancy formation energy significantly, by 35-40 $\%$ compared to
sum of individual doping.} This is likely
to improve the thermal stability at high temperature condition.

\textcolor{black}{Our computational exercise based on exhaustive search of experimental database, should
motivate future experimental processes in making high-performance 2:17 interstitial magnets, with
cheapest RE element Ce, the most abundant $3d$ metal, Fe and cheap non-metal interstitial dopings
like C and N. The estimated price-to-performance based on calculated energy product, and available
market price\cite{marketprice} turns out to be 0.03-0.22 USD/J. The enhanced thermal stability
of the carbo-nitrides compounds against the vacancy formation of the light elements further boosts
the promises of the suggested compounds.}

\section{Acknowledgement}
The authors acknowledge the support of DST Nano-mission for the computational
facility used in this study.

\section{Appendices}

\subsection{DFT details}

DFT calculations for electronic structure, magnetocrystalline anisotropy
are performed using the all-electron density-functional-theory code in
full potential linear augmented plane wave (FP-LAPW) basis, as implemented
in WIEN2K code.\cite{wien2k} For expensive structural optimization calculations, the plane wave based calculations, as implemented in Vienna Ab-initio Simulation Package (VASP),\cite{vasp} are carried out. The exchange-correlation functional is chosen to be
generalized-gradient approximation (GGA) of Perdew, Burke, and Ernzerhof.\cite{PBE} The localized nature of 4$f$ states of Ce
is captured through GGA+$U$ calculations,\cite{gga+u} with choice of $U$ = 6 eV and J$_{H}$ = 0.8 eV.
For light rare earths like Ce the $U$ value was shown to range from 4 eV to 7 eV, without affecting much
the physical properties.\cite{eriksson} The spin-orbit coupling effect at Ce, and TM sites are captured through
GGA+$U$+SOC calculations.\\
\indent
For FP-LAPW calculations, APW $+$ lo is used as the basis set, and the spherical harmonics are expanded upto $l =$ 10 and the charge density and potentials are represented upto $l =$ 6. The sphere radii are set at 2.5, 1.9, 2.34, 1.56 and 1.51 bohr for Ce, Fe, Co, N, and C. For good
convergence, a RK$_{max}$ value (the product of the smallest sphere radius and the largest plane-wave expansion wave vector) of
7.0 is used. We set the cutoff between core and valence states at $-$8.0 Ry. The k-space integrations are performed
with 112 k-points in irreducible Brillouin zone (BZ), following the report of use of 80 k-points in irreducible BZ in case of SmCo$_5$ to provide good estimate of MAE.\cite{mazin} Nevertheless, the convergence of results on k-space mesh is checked by carrying out calculation with 260 k-points.\\ 
\indent
The structural optimization in plane wave basis is carried out starting with experimental structure of Sm$_2$Fe$_{17}$CN, \cite{kou-cs} replacing
Sm with Ce, and relaxing all the internal coordinates until forces on all of the atoms become less than 0.001 eV/\AA. 
Upon moving from Sm 2:17 carbide/nitride interstitial compounds to Ce counterpart, the cell volume changes only nominally by
0.2$\%$ to 0.4$\%$.\cite{Pandey} For the plane wave calculations, energy cut-off of 600 eV and Monkhorst pack $k$-points
mesh of $8\times 8\times 8$ are used.\\
\indent
All the calculations are performed by considering a collinear spin arrangement. The MAE is obtained by
calculating the GGA+$U$+SOC total energies of the system, in FP-LAPW basis as \textcolor{black}{K$_u$} = E$_{a}$ - E$_{c}$ , where E$_a$ and E$_c$ are
the energies for the magnetization oriented along the crystallographic $a$ and $c$ directions, respectively. \textcolor{black}{For
  accurate estimates of vacancy formation energy, we also use FP-LAPW basis.}

\subsection{\textcolor{black}{Data preprocessing in Machine Learning}}

While constructing the database, we avoid inclusion of noisy data. We do bootstrapping to normalize the data which is followed by removal of outliers with the help of violin plot. A data is removed if it {lies outside of Q1-1.5$\times$IQR or Q3+1.5$\times$IQR, where IQR is the interquartile range and Q1, Q2 and Q3 are lower, median and upper quartile respectively.} In the next step we identify correlated attributes using Pearson's correlation coefficient which can be defined as, 

$$r=\dfrac{\sum_{i=1}^{i=n}(x_{i}-\bar{x})(y_{i}-\bar{y})}{\sqrt{\sum_{i=1}^{i=n}(x_{i}-\bar{x})^{2}}\sqrt{\sum_{i=1}^{i=n}(y_{i}-\bar{y})^{2}}}$$

Here $n$ is the sample size, $x_{i}$ and $y_{i}$ are sample points and $\bar{x}$ and $\bar{y}$ are the
sample means. 

\indent 
\begin{figure}[h]
\includegraphics[width=1.0\linewidth]{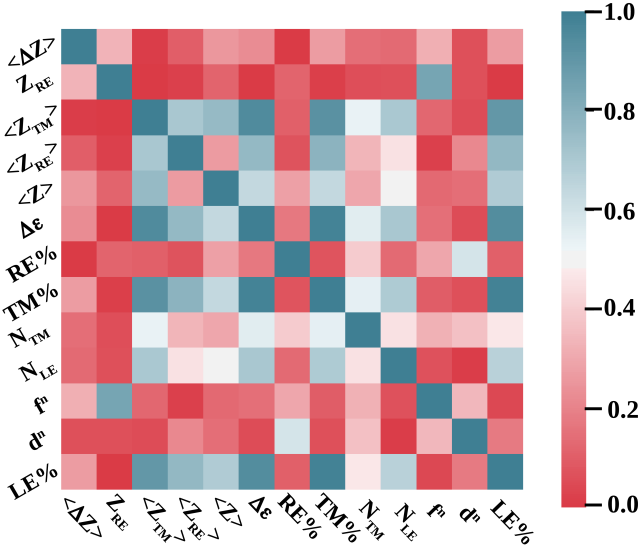}
\caption{(Color online) Heatmap indicating the correlation between different attributes considered to built ML algorithm. The color code is shown in the side panel. The boxes with red represent weak or no correlation, whereas blue boxes represent strong correlation between the attributes. }
\end{figure}

The heatmap obtained by using the above mentioned correlation is shown in Fig. 9. The correlation between the attributes is mapped between 0 and 1, considering the absolute values. The highly correlated attributes with correlation greater than 0.75 are as follows:
 \begin{enumerate}
 \item Electronegativity difference between RE and TM ($\Delta \epsilon$) and CW average of atomic no. of TM ($<Z_{TM}>$)
 \item CW TM percentage ($TM\%$) and CW average of atomic no. of TM ($<Z_{TM}>$).
 \item CW TM percentage ($TM\%$) and Electronegativity difference between RE and TM ($\Delta \epsilon$).
 \item Total number of f electrons ($f^n$) and Atomic no. of RE ($Z_{RE}$).
 \item  LE percentage ($LE\%$) and CW average of atomic no. of TM ($<Z_{TM}>$).
  \item  LE percentage ($LE\%$) and Electronegativity difference between RE and TM ($\Delta \epsilon$).
  \item  LE percentage ($LE\%$) and CW TM percentage ($TM\%$).
 \end{enumerate}
We thus discard $\Delta \epsilon$, $LE\%$, $Z_{RE}$ and $<Z_{TM}>$ from the list of attributes. 

\subsection{\textcolor{black}{Model construction for training in ML}}

\begin{figure}[h]
\includegraphics[width=1.0\linewidth]{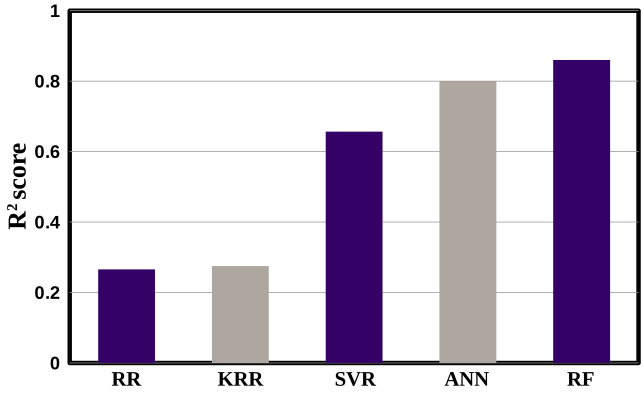}
\caption{(Color online) Coefficient of determination $R^2$ score of five different ML algorithms applied to T$_c$ dataset.}
\end{figure}

\begin{figure*}[]
\includegraphics[width=0.8\textwidth]{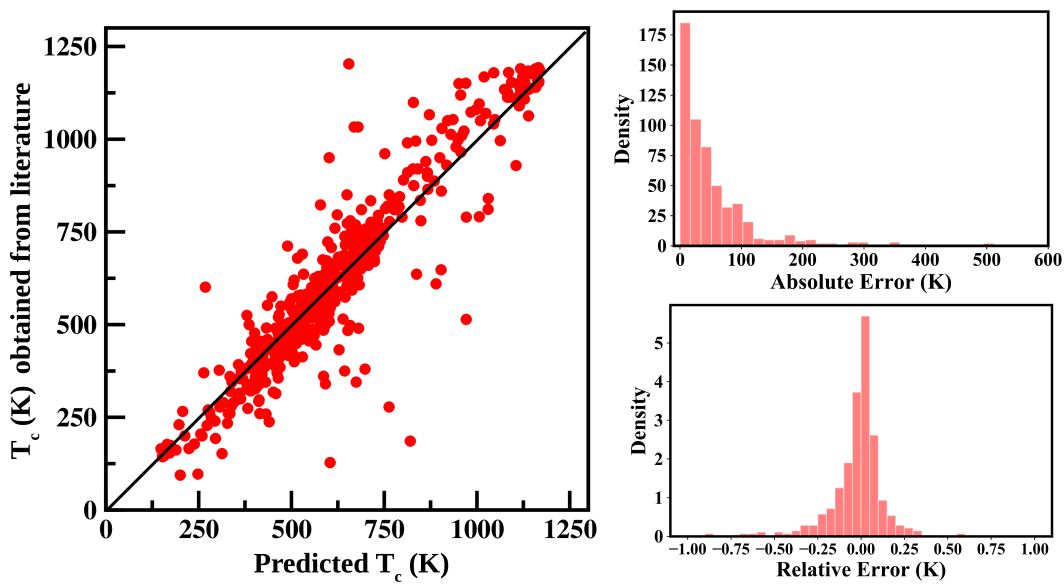}
\caption{(Color online) Model output from RF algorithm for T$_c$ of RE-TM intermetallics. The left panel shows the comparison of T$_c$ obtained from literature and predicted T$_c$.
 The distribution of absolute error between predicted T$_c$ and actual T$_c$ is shown in the
upper, right panel, while the lower, right panel presents the distribution of relative error for the compounds with T$_c > $ 600 K.}
\end{figure*}


The performance of a model can be quantified in terms of coefficient of determination which can be expressed as follows:\cite{R2}}
$$R^{2}=1-\dfrac{\sum_{i=1}^{N}[y_{i}-f(x_{i})]^{2}}{\sum_{i=1}^{N}[y_{i}-\mu]^{2}}$$
for predictions ${f(x_i)}$ and a set of actual values ${y_i}$ with mean $\mu$.
If the algorithm performs perfectly, $R^2$ score is 1. Fig. 10 shows score $R^2$ for five different algorithms.
\indent
 RR algorithm circumvents issues in ordinary linear regression like over-fitting or failure in finding unique solution due to multicollinearity. It develops on least square error by adding an extra penalty/regularization term to the loss function of ordinary linear regression. KRR builds on the ridge regression technique by using kernel {trick \cite{trick}} so that it can capture the nonlinearity present in the feature space. It can fit a non linear function by learning from a linear function spanned by a kernel which in turn mimics a non-linear function in the original space. SVR originated from support vector machines which are mainly popular in classification problem. It is based on the idea to search a hyperplane \cite{plane} by minimizing the error which is able to separate two different classes. SVR also uses kernel trick to map the data into a high dimensional feature space and then performs linear regression to fit the data. These three models are based on the same principle of linear regression and SVR is the best form according to our result. $R^2$ score is 0.66 for SVR whereas it is found to be poor ($\approx$ 0.25) for other two algorithms.

 \indent
 Apart from these we use two other algorithms{,} ANN and RF. The model performance scores are satisfactory for both of them. A simple ANN architecture called perceptron implements a processing element or artificial neuron called Threshold Logic Unit (TLU) which can have one or more input(s) and one output. Each input is related to a weight. The TLU calculates the weighted sum of its inputs, applies a step function (generally Heaviside or sign function) to it and outputs the result. A perceptron \cite{perceptron} is simply a layer of  TLUs operating in parallel and connected to all the inputs. Training an ANN model is equivalent to learning each weight factor in an iterative cycle. A more complex system (Multi-Layer Perceptron) can be built by associating additional interconnected layers to the architecture. A well functioning system consists of an input layer, several hidden layers and an output layer. In our case we have one input layer, two hidden layers where rectified linear unit (ReLU)\cite{relu} is used as activation function along with L2 regularization in the kernel, and an output layer. The constructed ANN model shows 0.80 as $R^2$ score. \\ 
 \indent
 Random forest is an ensemble method which consists of multiple decision trees. Each tree is built on a portion of entire training data with a subset of total number of attributes. Tree algorithm is based on 'top to bottom' approach, starting from a root node, it consists of many intermediate nodes and ends at leaf nodes. At each node of a tree a particular attribute classifies the data and helps to grow the tree. The prediction is based on accumulating the results from all such trees, taking ensemble average in case of regression or considering votes from majority trees in case of classification. Such an algorithm can capture the complex and nonlinear interaction between different attributes and can built a robust and sophisticated model.  Our random forest consists of 100 trees built by {bootstrapped\cite{boot}} sampling of the training set. Each tree allows checking a maximum of $\log_2$(number of features) while detecting the best split node. The quality of such a split is measured by using mean squared error (Gini index) in regression (classification). The model efficiency is calculated by running out-of-bag samples down each of the trees. We use ten-fold cross validation to extract the hyper-parameter and to construct the best model.\\ 
 \indent
 Fig. 11 shows the result of the best regression model using RF algorithm in case of T$_c$. The plot in the left panel shows the predicted T$_c$ versus T$_c$ obtained from experiments. The determination score $R^2$ is high enough (0.86), indicating a good agreement between the predicted T$_c$ and experimentally reported T$_c$. The mean absolute error in this model is 60 K. Additionally we evaluate absolute error and relative error for the compounds with T$_c >$ 600 K (cf Fig. 11, right panel). This analysis helps to determine the model performance for the compounds with T$_c >$ 600 K as we are interested to predict new RE-TM intermetallics with high T$_c$. The distribution of absolute error shows that for the most of the compounds ($\approx$ 85$\%$) the absolute error is less than 100 K. For 65$\%$ of the predicted cases, the absolute error is less than 50 K. We also check the absolute error for the compounds with T$_c <$ 300 K
 (not included in the figure). In this case our model predicts $\approx$ 76$\%$ compounds with absolute error less than 100 K and 50$\%$ instances are predicted with absolute error
 of 50 K. This observation prompts us to conclude that though the model prediction is in general good, it is less accurate for low T$_c$ compounds compared to high T$_c$ compounds. The distribution of relative error, expressed as $\epsilon_{rel}$ = (T$_c^{exp}$ -T$_c^{predicted}$)/T$_c^{exp}$, provides further support to this statement, which is shown in bottom, right panel of Fig. 11. The relative error distribution appears Gaussian like with slight asymmetry about the mean position. The relative error is less in the right side of the mean position than the left side suggesting the prediction of T$_c$ suffers less overestimation than underestimation. As found, only 1$\%$ of the instances are having $\epsilon_{rel}$ $>$ 50$\%$, 3$\%$ of the instances have 50$\%$ $>$ $\epsilon_{rel}$ $>$ 30$\%$ and 2$\%$ instances have 30$\% >$ $\epsilon_{rel} >$ 25$\%$, most
cases having tiny values of $\epsilon_{rel}$. This gives us confidence in accuracy of the predicted T$_c$ for compounds with T$_c$s exceeding 600 K.

\begin{figure}[h]
\includegraphics[width=1.0\linewidth]{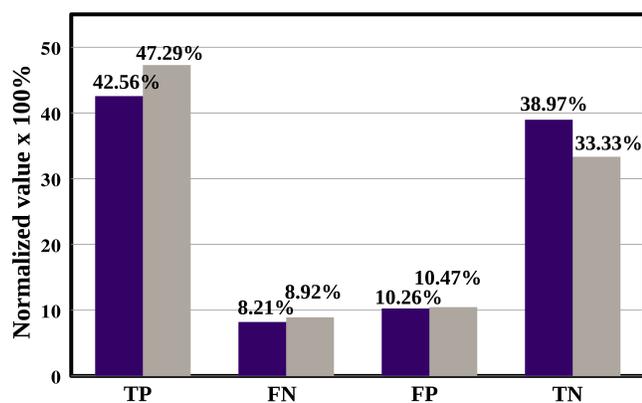}
\caption{(Color online) Normalized confusion matrix for $\mu_0$M$_s$(violet) and \textcolor{black}{K$_u$}(grey) classification using 10-fold cross-validation. Here positive (negative) class
 represents either compounds with $\mu_0$M$_s > (<)$ 1T, or
 compounds with uniaxial anisotropy i.e K$_u > (<)$ 0 MJ/m$^3$. True positive/negative or TP/TN are the compounds where their classes are predicted correctly. Whereas false positive (FP) and false negative (FN) are the off-diagonal terms of the matrix where the classes are incorrectly classified.}
\end{figure}
\indent
Turning to M$_s$, we use random forest algorithm to classify high M$_s$ from low M$_s$ compounds. The best model by performing 10-fold cross validation is built up with 81.53$\%$ accuracy. The resultant confusion matrix is shown in Fig. 12. For classification problem, F1 score determines the balance between precision and recall. In this case F1 score 82.2$\%$ indicates good anticipation with slight favour towards the prediction of compounds with high M$_s$ ($\mu_0$M$_s$ $>$ 1) (83.8$\%$) compared to the compounds with low M$_s$
($\mu_0$M$_s$ $<$ 1) (79.2$\%$). \\
\indent
Similar to M$_s$, we use random forest algorithm for \textcolor{black}{K$_u$}, to classify positive \textcolor{black}{K$_u$} from negative \textcolor{black}{K$_u$} compounds.
The best model by performing 10-fold cross validation, in this case, is built up with 80.62$\%$ accuracy
Like M$_s$, in this case F1 score for positive \textcolor{black}{K$_u$} is 83$\%$ and for negative is \textcolor{black}{K$_u$} 77.5$\%$ suggesting slight preference of classification towards positive \textcolor{black}{K$_u$} which is also captured in the plot of confusion matrix as shown in Fig. 12. \\
\indent

\end{document}